\documentclass[11pt,a4paper]{article}
\usepackage{amssymb}
\usepackage[dvips]{graphicx}
\usepackage{bm}
\usepackage{amsmath, textcomp}

\begin{document}

\title{On-shell renormalization scheme for ${\cal N}=1$ SQED and the NSVZ relation}

\author{
A.L.Kataev,\\
{\small{\em Institute for Nuclear Research of the Russian Academy of Sciences,}}\\
{\small{\em 117312, Moscow, Russia}};\\
{\small{\em Moscow Institute of Physics and Technology,}}\\
{\small{\em 141700, Dolgoprudny, Moscow Region, Russia}},\\
\\
A.E.Kazantsev\\
{\small{\em Moscow State University,}}\\
{\small{\em Faculty of Physics, Department of Quantum Theory and High Energy Physics,}}\\
{\small{\em 119991, Moscow, Russia}},\\
\\
K.V.Stepanyantz\\
{\small{\em Moscow State University,}}\\
{\small{\em Faculty of Physics, Department of Theoretical Physics,}}\\
{\small{\em 119991, Moscow, Russia}}}

\maketitle

\begin{abstract}
In this paper we investigate the renormalization of ${\cal N}=1$ supersymmetric quantum electrodynamics, regularized by higher derivatives, in the on-shell scheme. It is demonstrated that in this scheme the exact Novikov, Shifman, Vainshtein, and Zakharov (NSVZ) equation relating the $\beta$-function to the anomalous dimension of the matter superfields is valid in all orders of the perturbation theory. This implies that the on-shell scheme enters the recently constructed continuous set of NSVZ subtraction schemes. To verify this statement, we compare the anomalous dimension of the matter superfields in the two-loop approximation and the $\beta$-function in the three-loop approximation, which are explicitly calculated in this scheme. The finite renormalizations relating the on-shell scheme to some other NSVZ subtraction schemes formulated previously are obtained.
\end{abstract}

\unitlength=1cm

\vspace*{-18.0cm}

\begin{flushright}
INR-TH-2019-004
\end{flushright}

\vspace*{16.5cm}

\section{Introduction}
\hspace*{\parindent}

Among various renormalization schemes that can be used in quantum electrodynamics the subtraction on the mass shell is one of the most important (for a review, see, e.g., \cite{Grozin:2005yg}). The reason for this is that in this scheme renormalized quantities such as masses, charges, and anomalous magnetic moments can be subjected to direct experimental measurements. This distinguishes it from the (modified) minimal subtraction or momentum subtraction schemes.

The ${\cal N}=1$ supersymmetric generalization of quantum electrodynamics, besides the electron and photon, contains their superpartners, namely, a pair of complex scalar fields and a Majorana spinor. It is most convenient to describe this theory using the superfield formalism with the gauge fixing term respecting supersymmetry. In this case ${\cal N}=1$ supersymmetry appears to be a manifest symmetry of the theory, so that the perturbative calculations can be done in an ${\cal N}=1$ supersymmetric way. This is to be contrasted with the approach, when the gauge superfield is put into the Wess--Zumino gauge, in which only its physical components survive. Although in this case the quantization is performed in terms of physical fields only, the manifest supersymmetry is lost.

An important feature of ${\cal N}=1$ supersymmetric gauge theories is the existence of the relation between the $\beta$-function and the anomalous dimensions of the matter superfields \cite{Novikov:1983uc,Jones:1983ip,Novikov:1985rd,Shifman:1986zi}. For ${\cal N}=1$ supersymmetric electrodynamics (SQED) considered in this paper it can be written as \cite{Vainshtein:1986ja,Shifman:1985fi}

\begin{equation}\label{NSVZ}
\beta(\alpha)=\frac{N_f\alpha^2}{\pi}\Big(1-\gamma(\alpha)\Big).
\end{equation}

It is known that the NSVZ relation does not in general hold for an arbitrary renormalization prescription\footnote{The general equations describing how the NSVZ relation changes under finite renormalizations can be found in \cite{Kutasov:2004xu,Kataev:2014gxa}.} and is valid only in certain subtraction schemes called the NSVZ schemes.  Recently, it was discovered that all NSVZ subtraction schemes in ${\cal N}=1$ SQED form a class that can be parameterized by a single function and a single constant \cite{Goriachuk:2018cac}. Various schemes of this class are related by finite renormalizations satisfying a certain condition, which form a subgroup of the general renormalization group transformations \cite{Stueckelberg:1953dz,GellMann:1954fq,Bogolyubov:1956gh,Bogolyubov:1980nc}. This class in particular includes the so-called HD+MSL renormalization prescription (short for Higher Derivatives plus Minimal Subtraction of Logarithms) \cite{Kataev:2013eta,Kataev:2013csa}. In this case a theory is regularized by the higher covariant derivative method \cite{Slavnov:1971aw,Slavnov:1972sq} (see also Refs. \cite{Krivoshchekov:1978xg,West:1985jx,Aleshin:2016yvj} for its various ${\cal N}=1$ supersymmetric versions) and only powers of $\ln\Lambda/\mu$ are included into the renormalization constants. Here $\Lambda$ is the dimensionful parameter of the regularized theory, and $\mu$ is the subtraction point. Note the HD+MSL prescription gives the NSVZ- and NSVZ-like schemes for various theories, e.g., for the photino mass in the electrodynamics with softly broken supersymmetry \cite{Nartsev:2016mvn}\footnote{The NSVZ-like equation describing the renormalization of the gaugino mass has been proposed in \cite{Hisano:1997ua,Jack:1997pa,Avdeev:1997vx}.} or for the Adler $D$-function in ${\cal N}=1$ SQCD \cite{Kataev:2017qvk}.\footnote{This follows from the fact that the renormalization group functions defined in terms of the bare couplings satisfy the NSVZ and NSVZ-like equation with the higher derivative regularization. At present, it has been rigorously proved in all orders for ${\cal N}=1$ SQED in \cite{Stepanyantz:2011jy,Stepanyantz:2014ima}, for the renormalization of the photino mass in softly broken SQED in \cite{Nartsev:2016nym}, and for the Adler $D$-function in ${\cal N}=1$ SQCD in \cite{Shifman:2014cya,Shifman:2015doa}.} There are indications that the NSVZ scheme in the non-Abelian case is also given by this prescription in all orders  \cite{Stepanyantz:2016gtk}. This conjecture has been confirmed by explicit three-loop calculations in \cite{Shakhmanov:2017soc,Kazantsev:2018nbl}.

The $\overline{\mbox{DR}}$ scheme, most frequently used for practical calculations, does not enter the class of the NSVZ schemes, as was explicitly demonstrated in the three- \cite{Jack:1996vg} and four-loop \cite{Harlander:2006xq} approximations. Nevertheless, a finite renormalization of the coupling constant, specially tuned in each order of the perturbation theory, allows constructing the NSVZ scheme with dimensional reduction \cite{Jack:1996vg,Jack:1996cn,Jack:1998uj}. The difference between calculations with the higher derivative regularization and with dimensional reduction for ${\cal N}=1$ SQED in the three-loop approximation has been analyzed in Ref. \cite{Aleshin:2016rrr}.

Because the subtraction on the mass shell occupies a special place in electrodynamics, it would be interesting to find out whether a relation (\ref{NSVZ}) is satisfied in this scheme and, therefore, whether it falls into the class of NSVZ schemes. Using the results of Ref. \cite{Smilga:2004zr}, the guess was made that the NSVZ relation in ${\cal N}=1$ SQED is valid in the on-shell scheme \cite{Smilga_Private}. Note that the explicit calculations in Ref. \cite{Smilga:2004zr} were done only in the approximation, where the scheme dependence is not essential. In this paper we demonstrate that the NSVZ equation relating the $\beta$-function to the mass anomalous dimension is valid in the on-shell scheme in all orders. This statement is verified by the explicit calculation. Namely, the three-loop $\beta$-function is compared with the two-loop mass anomalous dimension in the on-shell scheme. This allows to check that Eq. (\ref{NSVZ}) really holds in this case.

\section{${\cal N}=1$ SQED: action and the higher derivative regularization}
\hspace*{\parindent}

In the superfield language $\mathcal{N}=1$ SQED with $N_{f}$ flavors of massive Dirac fermions and their superpartners is described by the action

\begin{eqnarray}
&& S = \frac{1}{4e_0^2} \mbox{Re} \int d^4x\, d^2\theta\, W^{a} W_{a} + \frac{1}{4} \sum\limits_{i=1}^{N_f}\int d^4x\, d^4\theta\, \Big(\phi_i^* e^{2V} \phi_i+\widetilde \phi_i^* e^{-2V} \widetilde{\phi_i}\Big)\nonumber\\
&& + \frac{1}{2} \sum\limits_{i=1}^{N_f} \Big(\int d^4x\, d^2\theta\, m_0\, \widetilde{\phi_i}\, \phi_i + \mbox{c.c.} \Big),
\end{eqnarray}

\noindent where $m_0$ is the bare mass of the chiral matter superfields. For simplicity, and in order not to deal with multiple thresholds, we assume the masses for different flavors to be equal.

The regularization is introduced by adding to the action the term with higher derivatives

\begin{equation}\label{SLambda}
S_{\Lambda} = \frac{1}{4e_0^2} \mbox{Re} \int d^4x\, d^2\theta\, W^{a} \Big[ R\Big(\frac{\partial^2}{\Lambda^2}\Big) - 1 \Big] W_{a},
\end{equation}

\noindent
where $R$ is a function which rapidly increases at large values of the argument and satisfies the condition $R(0)=1$. Moreover, to regularize divergences in the one-loop approximation, it is necessary to insert the Pauli--Villars determinants in the generating functional \cite{Slavnov:1977zf}. Following Ref. \cite{Kazantsev:2014yna}, let us introduce $n$ sets of the chiral Pauli--Villars superfields $\Phi_{iI}$ with masses $M_I$, where $I=1,\ldots, n$, and include

\begin{equation}\label{SPV}
S_{\mbox{\scriptsize PV}}=\sum\limits_{I=1}^{n}\left(\frac{1}{4}\sum\limits_{i=1}^{N_f}\int d^4x\, d^4\theta\, \Big(\Phi_i^*e^{2V}\Phi_i + \widetilde{\Phi}_i^* e^{-2V} \widetilde{\Phi}_i\Big) + \frac{1}{2}\sum\limits_{i=1}^{N_f}\Big(\int d^4x\, d^2\theta\, M \widetilde{\Phi}_i\Phi_i + \mbox{c.c.}\Big)\right)_I
\end{equation}

\noindent
into the total action. Then, to cancel one-loop divergences, their Grassmannian parities $(-1)^{P_{I}}$ and masses $M_I$ should satisfy the relations

\begin{equation}\label{Conditions_For_Coefficients}
\sum\limits_{I=1}^{n}(-1)^{P_I}+1=0;\qquad\sum\limits_{I=1}^{n}(-1)^{P_{I}}M_I^2+m_0^2=0.
\end{equation}

\noindent
In the massless case the masses of the Pauli--Villars superfields $M_I$ should be chosen proportional to the parameter $\Lambda$ in the higher derivative term. However, in the massive case it is convenient to present them in the form

\begin{equation}
M_I^2 = a_I^2 \Lambda^2 + b_I^2m_0^2,
\end{equation}

\noindent
where the coefficients $a_I$ and $b_I$, independent of the coupling constant, satisfy the equations

\begin{equation}
\sum\limits_{I=1}^n(-1)^{P_I}a_I^2=0; \qquad \sum\limits_{I=1}^n(-1)^{P_I}b_I^2+1=0,
\end{equation}

\noindent
which follow from Eq. (\ref{Conditions_For_Coefficients}). It should be noted that the derivative of $M_I/\Lambda$ with respect to $\ln \Lambda$ or $\ln m_0$ is of the order $m_0^2/\Lambda^2$ and, therefore, can be neglected in the limit $\Lambda\to\infty$.

To complete the quantization, the gauge-fixing term

\begin{equation}
S_{\mbox{\scriptsize gf}} = - \frac{1}{32 e_0^2 \xi_0} \int d^4x\, d^4\theta\, D^2 V R\Big(\frac{\partial^2}{\Lambda^2}\Big) \bar{D}^2V
\end{equation}

\noindent
is added to the action. Below we will use the Feynman gauge in which the renormalized gauge fixing parameter is fixed as $\xi=1$.

\section{The on-shell subtraction scheme}
\hspace*{\parindent}

To construct the on-shell scheme for ${\cal N}=1$ SQED, let us consider the part of the effective action quadratic in the matter superfields. It can be presented in the form

\begin{eqnarray}\label{Superfield_Action}
&&\Gamma^{(2)}_\phi=\frac{1}{4}\sum\limits_{i=1}^{N_f}\int \frac{d^4p}{(2\pi)^4}\, d^4\theta\, \Big(\phi_i^*(-p,\theta)\, \phi_i(p,\theta) + \widetilde \phi_i^*(-p,\theta)\, \widetilde{\phi_i}(p,\theta)\Big)\, G(p/\Lambda,m_0/\Lambda, \alpha_0)\qquad\nonumber\\
&& + \frac{1}{2}\sum\limits_{i=1}^{N_f}\Big(\int \frac{d^4p}{(2\pi)^4}\, d^2\theta\, m_0\, \widetilde{\phi_i}(-p,\theta)\, \phi_i(p,\theta)\, J(p/\Lambda, m_0/\Lambda,\alpha_0) + \mbox{c.c.}\Big),
\end{eqnarray}

\noindent
where the functions $G$ and $J$ are normalized in such a way that in the tree approximation $G=1$  and $J=1$. From the expression (\ref{Superfield_Action}) it is possible to construct the exact superfield propagators for the matter superfields, see Ref. \cite{Stepanyantz:2014ima} for details. In the coordinate representation they are written as

\begin{eqnarray}\label{Superfield_Propagators}
&&\hspace*{-3mm}\qquad\qquad\quad \Big(\frac{\delta^2\Gamma}{\delta\phi_{i x} \delta\phi^*_{j y}}\Big)^{-1} = \Big(\frac{\delta^2\Gamma}{\delta\widetilde \phi_{i x} \delta\widetilde\phi_{j y}^*}\Big)^{-1} = -\frac{G D^2_y \bar D^2_x}{4\big(\partial^2 G^2 + m_0^2 J^2\big)}\delta^8_{xy} \delta_{ij};\qquad\nonumber\\
&&\hspace*{-3mm} \Big(\frac{\delta^2\Gamma}{\delta\phi_{i x} \delta\widetilde \phi_{j y}}\Big)^{-1} = -\frac{m_0 J \bar D^2}{\partial^2 G^2 + m_0^2 J^2} \delta^8_{xy} \delta_{ij};\qquad \Big(\frac{\delta^2\Gamma}{\delta\phi_{i x}^* \delta\widetilde \phi_{j y}^*}\Big)^{-1} = -\frac{m_0 J D^2}{\partial^2 G^2 + m_0^2 J^2} \delta^8_{xy} \delta_{ij}.\qquad
\end{eqnarray}

\noindent
In the momentum representation all these propagators contain the denominator

\begin{equation}
p^2 G^2(p) - m_0^2 J^2(p),
\end{equation}

\noindent
where all arguments of the functions $G$ and $J$ except for the momentum $p$ were omitted.

The renormalized mass in the on-shell scheme is defined as the pole of the propagators (\ref{Superfield_Propagators}),

\begin{equation}
m=m_0\,\frac{J(p)}{G(p)}\biggr|_{p^2=m^2}.
\end{equation}

\noindent
It is convenient to introduce the mass renormalization constant $Z_m\equiv m_0/m$, which in the scheme under consideration is given by the expression

\begin{equation}
Z_m =\frac{G(p)}{J(p)}\biggr|_{p^2=m^2}.
\end{equation}

\noindent
The matter superfield renormalization constant $Z$ in the on-shell scheme is given by the residue at this pole. For all propagators (\ref{Superfield_Propagators}) the result is the same,

\begin{equation}
Z^{-1}=G(p)\Big(1+2m^2\frac{\partial}{\partial p^2}\ln \frac{G(p)}{J(p)}\Big)\biggr|_{p^2=m^2}.
\end{equation}

Note that, due to the superpotential non-renormalization in ${\cal N}=1$ supersymmetric theories \cite{Grisaru:1979wc}, it is usually assumed that $Z Z_m =1$. However, in the one-shell scheme it is not so, because

\begin{equation}\label{Relation}
Z^{-1}Z_m^{-1}=J(p)\biggl(1+2m^2\frac{\partial}{\partial p^2}\ln \frac{G(p)}{J(p)}\biggr)\biggr|_{p^2=m^2}.
\end{equation}

\noindent
Although this expression is not equal to 1, it is finite in the ultraviolate  region due to the non-renormalization of the superpotential. This implies that the renormalization constants $Z$ and $Z_m^{-1}$ differ by a finite factor.

Note that in the component formulation of the theory the scalars and the spinors will have the same renormalization constants only if the theory is regularized and quantized in a manifestly supersymmetric way. In the case of using the Wess--Zumino gauge this equality will be lost. This can be seen already in the one-loop approximation, see Ref. \cite{Wess:1974jb}. On the other hand, since the relation between the bare and the pole mass must be gauge-independent, the equality between the fermion and the scalar masses must be preserved after renormalization in the on-shell scheme whichever of the two quantization methods is used \cite{Goity:1983aw}.

Quantum corrections to the two-point Green function of the gauge superfield are encoded in the function $d(k/\Lambda, m_0/\Lambda, \alpha_0)$, which enters the  effective action as

\begin{equation}
\Gamma^{(2)}_V - S_{\mbox{\scriptsize gf}} = -\frac{1}{16\pi} \int\frac{d^4k}{(2\pi)^4}\, d^4\theta\, V(-k,\theta) \partial^2\Pi_{1/2}V(k,\theta)\, d^{-1}(k/\Lambda, m_0/\Lambda, \alpha_0),
\end{equation}

\noindent
where $\partial^2\Pi_{1/2}\equiv - D^{a} \bar{D}^2 D_{a}/8$, and the normalization constant is chosen in such a way that in the tree approximation $d^{-1}=\alpha_0^{-1}$. The function $d(k/\Lambda, m_0/\Lambda,\alpha_0)$ is the invariant charge \cite{Bogolyubov:1980nc} of the supersymmetric quantum electrodynamics. In the limit $k\to 0$ it gives the value of the fine-structure constant as a function of $m_0/\Lambda$ and $\alpha_0$ in the supersymmetric case. In the on-shell subtraction scheme this value plays the role of the renormalized coupling constant $\alpha$. The $\beta$-function in this scheme is defined as

\begin{equation}
\beta(\alpha)=\frac{d\alpha}{d \ln m }\biggr|_{\alpha_0,\Lambda=\mbox{const}},
\end{equation}

\noindent
where $m$ is the pole mass defined earlier.

\section{The three-loop $\beta$-function in the on-shell scheme}
\hspace*{\parindent}

An important feature of using the higher covariant derivative regularization in supersymmetric theories is the factorization of the loop integrals contributing to the function $d^{-1}(k/\Lambda, m_0/\Lambda,\alpha_0)$ in the limit $k\to 0$ into integrals of double total derivatives with respect to the momenta. This was first discovered in explicit calculations for ${\cal N}=1$ SQED in \cite{Soloshenko:2003nc} (total derivatives) and \cite{Smilga:2004zr} (double total derivatives). The rigorous all-order proof for the Abelian case has been done in Refs. \cite{Stepanyantz:2011jy, Stepanyantz:2014ima}. (The factorization into double total derivatives seems to be a general feature of supersymmetric theories and theories with softly broken supersymmetry regularized by higher covariant derivatives, see, e.g., the calculations of Refs. \cite{Nartsev:2016nym,Shakhmanov:2017soc,Kazantsev:2018nbl,Pimenov:2009hv,Stepanyantz:2011bz,Buchbinder:2015eva}.)

In ${\cal N}=1$ SQED the double total derivatives are taken with respect to the momenta of the matter loops to which the external lines of the gauge superfield are attached. If a double total derivative acts on a massless propagator, it produces a delta-function singularity which gives rise to a nonvanishing contribution. However, if a double total derivative acts only on massive propagators, the integral of this total derivative vanishes. This implies that in massive ${\cal N}=1$ SQED the only nonvanishing contribution to the function $d^{-1}(k/\Lambda, m_0/\Lambda,\alpha_0)$ at $k=0$ comes from the one-loop approximation. In the case of using the higher derivative regularization it is possible to write the one-loop contribution to this function in the form\footnote{In our notation capital letters denote Euclidean momenta.}

\begin{eqnarray}\label{D_OneLoop}
&& d^{-1}(K/\Lambda=0, m_0/\Lambda,\alpha_0) = \alpha_0^{-1} + 2\pi N_f \int\frac{d^4Q}{(2\pi)^4} \frac{\partial}{\partial Q^\mu}\frac{\partial}{\partial Q_\mu}\nonumber\\
&&\qquad\qquad\qquad\qquad\qquad\qquad \times  \left(-\frac{1}{Q^2}\ln \big(Q^2+m_0^2\big) + \smash{\sum\limits_{I=1}^n} c_I \frac{1}{Q^2} \ln \big(Q^2+M_I^2\big)\right),\qquad
\end{eqnarray}

\noindent
where $c_I=(-1)^{P_I+1}$, see Refs. \cite{Soloshenko:2003nc,Stepanyantz:2012zz}. It is important that this expression is exact. All higher order contributions in the massive case vanish as integrals of total derivatives acting on non-singular functions \cite{Stepanyantz:2011jy,Stepanyantz:2014ima}. Note that the singularities are absent, because all propagators are massive.

The integral in Eq. (\ref{D_OneLoop}) can easily be calculated, see, e.g., \cite{Aleshin:2016yvj}. Taking into account that $d(0, m_0/\Lambda,\alpha_0)=\alpha$ is the renormalized charge in the on-shell scheme and omitting terms suppressed by powers of $m_0/\Lambda$, we obtain

\begin{equation}
\alpha^{-1}-\alpha_0^{-1}=\frac{N_f}{\pi}\Big(\ln\frac{\Lambda}{m_0}+\sum\limits_{I=1}^{n}c_I\ln a_I\Big).
\end{equation}

\noindent
Next, following \cite{Smilga:2004zr}, the right-hand side is expressed in terms of the renormalized mass,

\begin{equation}\label{ChargeRenorm}
\alpha^{-1}-\alpha_0^{-1}=\frac{N_f}{\pi}\Big(\ln\frac{\Lambda}{m}-\ln Z_m+\sum\limits_{I=1}^{n}c_I\ln a_I\Big).
\end{equation}

\noindent Then differentiating with respect to $\ln m$ gives the NSVZ relation

\begin{equation}\label{RelationSortOf}
\beta(\alpha)=\frac{N_f\alpha^2}{\pi}\Big(1+\gamma_m(\alpha)\Big)
\end{equation}

\noindent
written in terms of the mass anomalous dimension

\begin{equation}\label{TrueNSVZ}
\left.\gamma_{m}(\alpha)=\frac{d\ln Z_m}{d\ln m}\right|_{\alpha_0,\Lambda=\mbox{const}}.
\end{equation}

\noindent
Thus, the NSVZ equation similar to Eq. (\ref{NSVZ}) is indeed valid in the on-shell scheme. It relates the $\beta$-function in a given order to the mass anomalous dimension in the previous order. Note that in the on-shell scheme the mass anomalous dimension differs from the anomalous dimension of the matter superfields taken with the opposite sign.\footnote{Exactly as in the case of (non-supersymmetric) QED \cite{Broadhurst:1991fy}, in the lowest-order approximation the corresponding renormalization constants differ by an ultraviolet finite but infrared divergent term, $Z^{-1} Z_m^{-1} = 1 + \alpha (1-\ln m/\kappa)/\pi + O(\alpha^2)$, where $\kappa$ is a small photon mass.}

Using Eq. (\ref{RelationSortOf}) it is possible to construct the three-loop $\beta$-function in the on-shell scheme by calculating the mass anomalous dimension in the two-loop approximation. This is done in this paper for the theory regularized by higher derivatives. Methods of evaluating Feynman integrals with the help of this regularization are not described in the literature in enough detail, while there is some interest in investigating various $D=4$ techniques for calculating quantum corrections (see the review \cite{Gnendiger:2017pys}).  That is why in Appendix \ref{Appendix_Z} we describe in detail how the renormalization constant $Z_m$ is obtained in the two-loop approximation. The result is given by the expression

\begin{eqnarray}\label{MassRenorm}
&& \ln Z_m = -\frac{\alpha_0}{\pi} \biggl(\ln\frac{\Lambda}{m}+\frac{1}{2}-\frac{A}{2}\biggr)+\frac{\alpha_0^2}{\pi^2}\biggl(\frac{N_f}{2}\ln^2\frac{\Lambda}{m}
\nonumber\\
&&\qquad\qquad\qquad\qquad\qquad
+\ln\frac{\Lambda}{m} \Big( \frac{3N_f}{2} + \frac{1}{2} + N_f \sum\limits_{I=1}^{n}c_{I}\ln a_I\Big) + O(1)\biggr)+O(\alpha_0^3).\qquad
\end{eqnarray}

\noindent
In this equation the symbol $O(1)$ denotes finite terms that do not vanish in the limit $\Lambda\to\infty$, and the constant $A$ is defined by

\begin{equation}\label{A_Definition}
A\equiv 2\int\limits_{0}^{\infty}dK \ln\frac{K}{\Lambda}\, \frac{d}{dK}\Big(\frac{1}{R_K}\Big),
\end{equation}

\noindent
where $R_K\equiv R(K^2/\Lambda^2)$. (For the regulator $R(K^2/\Lambda^2)=1+(K^{2}/\Lambda^2)^n$ this integral vanishes, so that $A=0$.) Differentiating (\ref{MassRenorm}) with respect to $\ln m$ and expressing the result in terms of $\alpha$ using (\ref{ChargeRenorm}) we obtain

\begin{equation}\label{TwoLoopGamma}
\gamma_m(\alpha)=\frac{d\ln Z_m}{d\ln m}=\frac{\alpha}{\pi}-\frac{\alpha^2(3N_f+1)}{2\pi^2}+O(\alpha^3).
\end{equation}

\noindent
As expected in the on-shell scheme, any dependence on the regularization details has disappeared. After substituting the mass anomalous dimension (\ref{TwoLoopGamma}) into Eq. (\ref{RelationSortOf}) the three-loop result for the $\beta$-function takes the form

\begin{equation}\label{BetaInOS}
\beta(\alpha) = \frac{N_f\alpha^2}{\pi}\left(1 + \frac{\alpha}{\pi} - \frac{\alpha^2(3N_f+1)}{2\pi^2} + O(\alpha^3)\right).
\end{equation}

\noindent
Comparing it with the corresponding result in the $\overline{\mbox{DR}}$-scheme (i.e., in the case of using dimensional reduction supplemented by modified minimal subtractions) \cite{Jack:1996vg}

\begin{equation}\label{BetaInDred}
\beta_{\overline{\mbox{\scriptsize DR}}}(\alpha_{\overline{\mbox{\scriptsize DR}}})=\frac{\alpha_{\overline{\mbox{\scriptsize DR}}}^2 N_f}{\pi}\Bigl(1+\frac{\alpha_{\overline{\mbox{\scriptsize DR}}}}{\pi}-\frac{\alpha_{\overline{\mbox{\scriptsize DR}}}^2}{4\pi^2}(3N_f+2) + O(\alpha_{\overline{\mbox{\scriptsize DR}}}^3)\Bigr),
\end{equation}

\noindent
we see that the terms linear in $N_f$ coincide. This follows from the scheme-independence of these terms proved in \cite{Kataev:2013csa} in all orders, which is related to the so-called conformal symmetry limit of perturbative quenched quantum electrodynamics \cite{Kataev:2013vua}.

\section{Relations between the on-shell scheme and other NSVZ schemes}
\hspace*{\parindent}

In the previous section it was demonstrated that the NSVZ relation (\ref{RelationSortOf}) is valid in the on-shell scheme in all orders. Therefore, this scheme belongs to the class of NSVZ schemes described in Ref. \cite{Goriachuk:2018cac}, which also includes the all-order HD+MSL prescription and the NSVZ scheme constructed with dimensional reduction in the three-loop approximation in Refs. \cite{Jack:1996vg,Aleshin:2016rrr}. According to Ref. \cite{Goriachuk:2018cac} any two NSVZ subtraction schemes can be related by a finite renormalization

\begin{equation}\label{FiniteRenorm}
\alpha'(\alpha_0,\Lambda/\mu) = \alpha'(\alpha(\alpha_0,\Lambda/\mu)); \qquad  Z'(\alpha'(\alpha),\Lambda/\mu) = z(\alpha)\, Z(\alpha,\Lambda/\mu),
\end{equation}

\noindent which is subjected to the constraint

\begin{equation}\label{Constraint}
\frac{1}{\alpha'(\alpha)} - \frac{1}{\alpha} = \frac{N_f}{\pi}\ln z(\alpha) + B,
\end{equation}

\noindent where $B$ is a constant.

First, let us find the finite renormalization relating the on-shell scheme to the HD+MSL scheme. According to the HD+MSL prescription, the calculations are to be carried out with the higher derivative regularization and only powers of $\ln\Lambda/\mu$ are included into renormalization constants, so that in this scheme

\begin{eqnarray}
&& \frac{1}{\alpha_{\mbox{\scriptsize HD+MSL}}} = \frac{1}{\alpha_0} + \frac{N_f}{\pi} \ln\frac{\Lambda}{\mu} + \frac{\alpha_0 N_f}{\pi^2} \ln\frac{\Lambda}{\mu} + O(\alpha_0^2);\qquad\nonumber\\
&& \ln Z_{\mbox{\scriptsize HD+MSL}} = \frac{\alpha_0}{\pi} \ln\frac{\Lambda}{\mu} + O(\alpha_0^2).
\end{eqnarray}

\noindent
Comparing these equations with Eqs. (\ref{ChargeRenorm}) and (\ref{MassRenorm}) we derive the required finite renormalization,

\begin{eqnarray}\label{HDOS}
&& \alpha_{\mbox{\scriptsize HD+MSL}}^{-1}\Big|_{\mu=m} =\alpha_{\mbox{\scriptsize OS}}^{-1} - \frac{N_f}{\pi} \sum\limits_{I=1}^n c_I \ln a_I +\frac{N_f\alpha_{\mbox{\scriptsize OS}}}{\pi^2} \Big(-\frac{1}{2} + \frac{A}{2}\Big) + O(\alpha_{\mbox{\scriptsize OS}}^2);\quad\nonumber\\
&& \ln Z_{\mbox{\scriptsize HD+MSL}}\Big|_{\mu=m} +\ln Z_m = \frac{\alpha_{\mbox{\scriptsize OS}}}{\pi}\Big(-\frac{1}{2} + \frac{A}{2}\Big) + O(\alpha_{\mbox{\scriptsize OS}}^2).
\end{eqnarray}

\noindent
Evidently, in this case the condition \eqref{Constraint} is satisfied with $B = -(N_f/\pi)\sum\limits_{I=1}^n c_I \ln a_I$.

Also it is possible to find the finite renormalization which relates the on-shell scheme to the NSVZ scheme constructed with the dimensional reduction. (For short, we will call this scheme ``DR+NSVZ''.) In the case of using the DR+NSVZ scheme the renormalization group functions (RGFs) can be found in \cite{Aleshin:2016rrr} and have the form

\begin{eqnarray}
&&\hspace*{-6mm} \frac{\beta_{\mbox{\scriptsize DR+NSVZ}}\big(\alpha_{\mbox{\scriptsize DR+NSVZ}}\big)}{\alpha_{\mbox{\scriptsize DR+NSVZ}}^2 } = \frac{N_f}{\pi}\Big(1+\frac{\alpha_{\mbox{\scriptsize DR+NSVZ}}}{\pi} -\frac{\alpha_{\mbox{\scriptsize DR+NSVZ}}^2}{2\pi^2}\big(1+N_f\big) + O\big(\alpha_{\mbox{\scriptsize DR+NSVZ}}^3\big) \Big);\nonumber\\
&&\hspace*{-6mm} \gamma_{\mbox{\scriptsize DR+NSVZ}}\big(\alpha_{\mbox{\scriptsize DR+NSVZ}}\big) = -\frac{\alpha_{\mbox{\scriptsize DR+NSVZ}}}{\pi} + \frac{\alpha_{\mbox{\scriptsize DR+NSVZ}}^2}{2\pi^2}\big(1+N_f\big) + O\big(\alpha_{\mbox{\scriptsize DR+NSVZ}}^3\big).
\end{eqnarray}

\noindent
These expressions should be compared with Eqs. (\ref{TwoLoopGamma}) and (\ref{BetaInOS}). With the help of the standard equations describing how RGFs transform under finite renormalizations \cite{Vladimirov:1979my}
we obtain the finite renormalization after which RGFs in the on-shell scheme are converted into RGFs in the DR+NSVZ scheme,

\begin{eqnarray}\label{DROS}
&& \alpha_{\mbox{\scriptsize DR+NSVZ}}^{-1}\Big|_{\mu=m} =\alpha_{\mbox{\scriptsize OS}}^{-1} + \frac{N_f}{\pi} z_1 +\frac{N_f \alpha_{\mbox{\scriptsize OS}}}{\pi^2} \big(-1 + z_1\big) + O(\alpha_{\mbox{\scriptsize OS}}^2);\quad\nonumber\\
&& \ln Z_{\mbox{\scriptsize DR+NSVZ}}(\alpha_{\mbox{\scriptsize DR+NSVZ}})\Big|_{\mu=m} + \ln Z_{m,\, \mbox{\scriptsize OS+DRED}}(\alpha_{\mbox{\scriptsize OS}}) = \frac{\alpha_{\mbox{\scriptsize OS}}}{\pi}\big(-1 + z_1\big) + O(\alpha_{\mbox{\scriptsize OS}}^2).\qquad
\end{eqnarray}

\noindent
However, these equations contain an undefined constant $z_1$, which reflects the arbitrariness of choosing a renormalization point in the DR+NSVZ scheme. This constant can be found by comparing the one-loop expressions for the renormalized function $d^{-1}$ in the limit $k\to 0$,

\begin{equation}
\alpha_{\mbox{\scriptsize OS}}^{-1} = d^{-1}_{\mbox{\scriptsize OS}}\Big|_{k\to 0} = d^{-1}_{\mbox{\scriptsize DR+NSVZ}}\Big|_{k\to 0} = \alpha_{\mbox{\scriptsize DR+NSVZ}}^{-1} + \frac{N_f}{\pi} \ln\frac{\mu}{m} + O(\alpha_{\mbox{\scriptsize DR+NSVZ}}),
\end{equation}

\noindent
so that $z_1=0$. This is analogous to the case of (non-supersymmetric) QED in which a similar coefficient also vanishes, $\alpha^{-1}_{\overline{\mbox{\scriptsize MS}}}\Big|_{\mu=m} = \alpha^{-1}_{\mbox{\scriptsize OS}} + O(\alpha_{\mbox{\scriptsize OS}})$, see Ref. \cite{Broadhurst:1992za}.

In the case $\mu\ne m$ the considered finite renormalization takes the form

\begin{eqnarray}\label{DROS}
&& \alpha_{\mbox{\scriptsize DR+NSVZ}}^{-1} =\alpha_{\mbox{\scriptsize OS}}^{-1} - \frac{N_f}{\pi} \ln \frac{\mu}{m} -\frac{N_f \alpha_{\mbox{\scriptsize OS}}}{\pi^2} \big(1 + \ln \frac{\mu}{m}\big) + O(\alpha_{\mbox{\scriptsize OS}}^2);\quad\nonumber\\
&& \ln Z_{\mbox{\scriptsize DR+NSVZ}}(\alpha_{\mbox{\scriptsize DR+NSVZ}}) + \ln Z_{m,\, \mbox{\scriptsize OS+DRED}}(\alpha_{\mbox{\scriptsize OS}}) = -\frac{\alpha_{\mbox{\scriptsize OS}}}{\pi}\big(1 + \ln \frac{\mu}{m}\big) + O(\alpha_{\mbox{\scriptsize OS}}^2).\qquad
\end{eqnarray}

\noindent
One can easily verify that the constraint \eqref{Constraint} is also satisfied for the functions (\ref{DROS}) with $B=-N_f \ln(\mu/m)/\pi$.

\section{Conclusion}
\hspace*{\parindent}

We have explicitly demonstrated that the NSVZ equation in ${\cal N}=1$ SQED is valid in the on-shell scheme in all orders. In this case it relates the $\beta$-function to the mass anomalous dimension. The NSVZ relation appears in the on-shell scheme due to the fact that quantum corrections to the photon polarization operator in the limit of zero momentum are given by integrals of double total derivatives with the higher derivative regularization. In the massive case these total derivatives act on nonsingular expressions in all orders beyond the one-loop approximation. The remaining one-loop contribution produces the NSVZ relation between the $\beta$-function and the mass anomalous dimension. This implies that the $\beta$-function in a given order can be found by calculating the mass anomalous dimension in the previous order. In this paper, having calculated the latter to the two-loop order, we obtained the $\beta$-function in the on-shell scheme to the three-loop order.

It was also investigated how the on-shell scheme in ${\cal N}=1$ SQED is related to other known NSVZ schemes, namely HD+MSL and the NSVZ scheme based on dimensional reduction. Finite renormalizations relating the on-shell scheme to these two schemes have been constructed. They were shown to satisfy the constraint (\ref{Constraint}) derived in \cite{Goriachuk:2018cac}.

\section*{Acknowledgments}
\hspace*{\parindent}

The work of ALK and AEK was supported by the Foundation for the Advancement of Theoretical Physics and Mathematics 'BASIS', grant \textnumero 17-11-120.

\appendix

\section{The renormalization constant $Z_m$ in the two-loop order}
\hspace*{\parindent}\label{Appendix_Z}

This appendix is devoted to the calculation of the two-loop renormalization constant $Z_m$ in the on-shell scheme. In particular, we describe the technique of evaluating the $D=4$ loop integrals appearing with the higher derivative regularization.

\subsection{$Z_m$ as a sum of loop integrals}
\hspace*{\parindent}

In the considered approximation the logarithm of the renormalization constant $Z_m$ is written as

\begin{equation}\label{Zmass}
\ln Z_m =\ln G(m)-\ln J(m)=\Delta G(m)-\Delta J(m)-\frac{(\Delta G(m))^2}{2}+\frac{(\Delta J(m))^2}{2}+\ldots,
\end{equation}

\noindent
where $\Delta G \equiv G-1$ and $\Delta J \equiv J-1$. Using the results of Ref. \cite{Kazantsev:2014yna},\footnote{With the higher derivative regularization the superdiagrams contributing to the function $G$ have first been calculated in Refs. \cite{Soloshenko:2002np,Soloshenko:2003sx}.} after the Wick rotation it is possible to present this expression as a sum of Euclidean loop integrals

\begin{equation}\label{I_One-Loop}
\ln Z_m = \alpha_0 I_{\mbox{\scriptsize one-loop}} + \alpha_0^2\Big(I_1 + I_2 + I_3 + I_4 + I_5 -\frac{1}{2} \big(I_{\mbox{\scriptsize one-loop}}\big)^2\Big) + O(\alpha_0^3).
\end{equation}

\noindent
In explicit expressions for these integrals (presented below), Euclidean momenta will be denoted by capital letters. Due to the higher derivative regularization, denominators of the integrands contain the function $R_K$. In the simplest case it can be chosen as $R_K = 1+K^{2n}/\Lambda^{2n}$. However, in the general case considered here it is sufficient to require that $R_K(0)=1$ and (due to the presence of higher powers of the momentum) $R_K\to \infty$ in the limit $K\to \infty$. In Eq. (\ref{I_One-Loop})

\begin{equation}
\left.\alpha_0 I_{\mbox{\scriptsize one-loop}} = - 8\pi\alpha_0 \int \frac{d^4K}{(2\pi)^4} \frac{1}{K^2 R_K \big((P+K)^2 + m^2\big)}\right|_{P^2=-m^2}
\end{equation}

\noindent
is the one-loop contribution, while the remaining integrals

\begin{eqnarray}\label{TheTotal}
&& \hspace*{-11mm} \left. I_1 \equiv   \frac{m_0^2-m^2}{\alpha_0} \int\frac{d^4K}{(2\pi)^4} \frac{8\pi}{K^2 R_K ((P+K)^2+m^2)^2}\right|_{P^2=-m^2};\\
&& \hspace*{-11mm} I_2 \equiv -128\pi^2 \int\frac{d^4K}{(2\pi)^4} \frac{d^4L}{(2\pi)^4} \frac{P_\mu (K+L)^\mu - m^2}{K^2 R_K L^2 R_L ((P+K)^2+m^2) ((P+L)^2+m^2)}\nonumber\\
&& \hspace*{-11mm} \qquad\qquad\qquad\qquad\qquad\qquad\qquad\qquad\qquad\qquad\qquad\quad\ \left. \times \frac{1}{((P+K+L)^2+m^2)} \right|_{P^2=-m^2};\\
&& \hspace*{-11mm} \left. I_3 \equiv 128\pi^2 \int\frac{d^4K}{(2\pi)^4} \frac{d^4L}{(2\pi)^4} \frac{m^2}{K^2 R_K L^2 R_L ((P+K)^2+m^2)^2 ((P+K+L)^2+m^2)}\right|_{P^2=-m^2};\\
&& \hspace*{-11mm} \left. I_4 \equiv 64\pi^2 \int\frac{d^4K}{(2\pi)^4} \frac{d^4L}{(2\pi)^4} \frac{1}{K^2 R_K L^2 R_L ((P+K)^2+m^2) ((P+K+L)^2+m^2)}\right|_{P^2=-m^2};\\
&& \hspace*{-11mm} I_5 \equiv 64\pi^2 N_f \int\frac{d^4K}{(2\pi)^4} \frac{d^4L}{(2\pi)^4} \frac{1}{K^2 R_K^2 ((P+K)^2+m^2)}\left(\frac{1}{(L^2+m^2)((L+K)^2+m^2)}\right.\nonumber\\
&& \hspace*{-11mm} \hspace*{-1mm}\qquad\qquad\qquad\qquad\qquad\qquad\qquad\quad\ \ \left.\left.\vphantom{\frac{1}{2}}\smash{ + \sum\limits_{I=1}^n(-1)^{P_I}\frac{1}{(L^2+M_I^2)((L+K)^2+M_I^2)}}\right)\right|_{P^2=-m^2}
\end{eqnarray}

\noindent
correspond to the two-loop approximation. Note that in these integrals terms proportional to $\alpha_0^2\,(P^2+m^2)$ were omitted, because they evidently vanish due to the condition $P^2=-m^2$. Also all these integrals were expressed in terms of the renormalized mass $m$. Therefore, the one-loop superdiagrams give both the integral $I_{\mbox{\scriptsize one-loop}}$ and the integral $I_1$. (The latter one is produced by the one-loop superdiagrams containing an insertion of the one-loop mass counterterm.)

\subsection{One-loop contribution}
\hspace*{\parindent}\label{Appendix_One-Loop}

The one-loop contribution is given by the expression

\begin{equation}\label{Z_M_One-Loop}
\left.\ln Z_m =\ln G(m)-\ln J(m)= -  \int \frac{d^4K}{(2\pi)^4} \frac{8\pi\alpha_0}{K^2 R_K \big((P+K)^2 + m^2\big)}\right|_{P^2=-m^2} + O(\alpha_0^2),
\end{equation}

\noindent
where the bare mass $m_0$ was replaced by the renormalized mass $m$, because their difference $\Delta m^2 \equiv m^2-m_0^2$ is proportional to $\alpha_0$ and is essential in the next order. The integral in Eq. (\ref{Z_M_One-Loop}) can be calculated in four-dimensional spherical coordinates using the method of Refs. \cite{Soloshenko:2002np,Soloshenko:2003sx}. Introducing the variable $x\equiv \cos\theta_3$, where $\theta_3$ is the angle between the vectors $K^\mu$ and $P^\mu$, it can be rewritten in the form

\begin{equation}\label{1Loop_Integral_With_C}
\ln Z_m = - \alpha_0 \int\limits_0^\infty dK\, \frac{1}{\pi^2 R_K} \oint\limits_{\cal C} dx\, \frac{\sqrt{1-x^2}}{K + 2im x} + O(\alpha_0^2).
\end{equation}

\begin{figure}[h]
\begin{picture}(0,6)
\put(5,0){\includegraphics[scale=0.3]{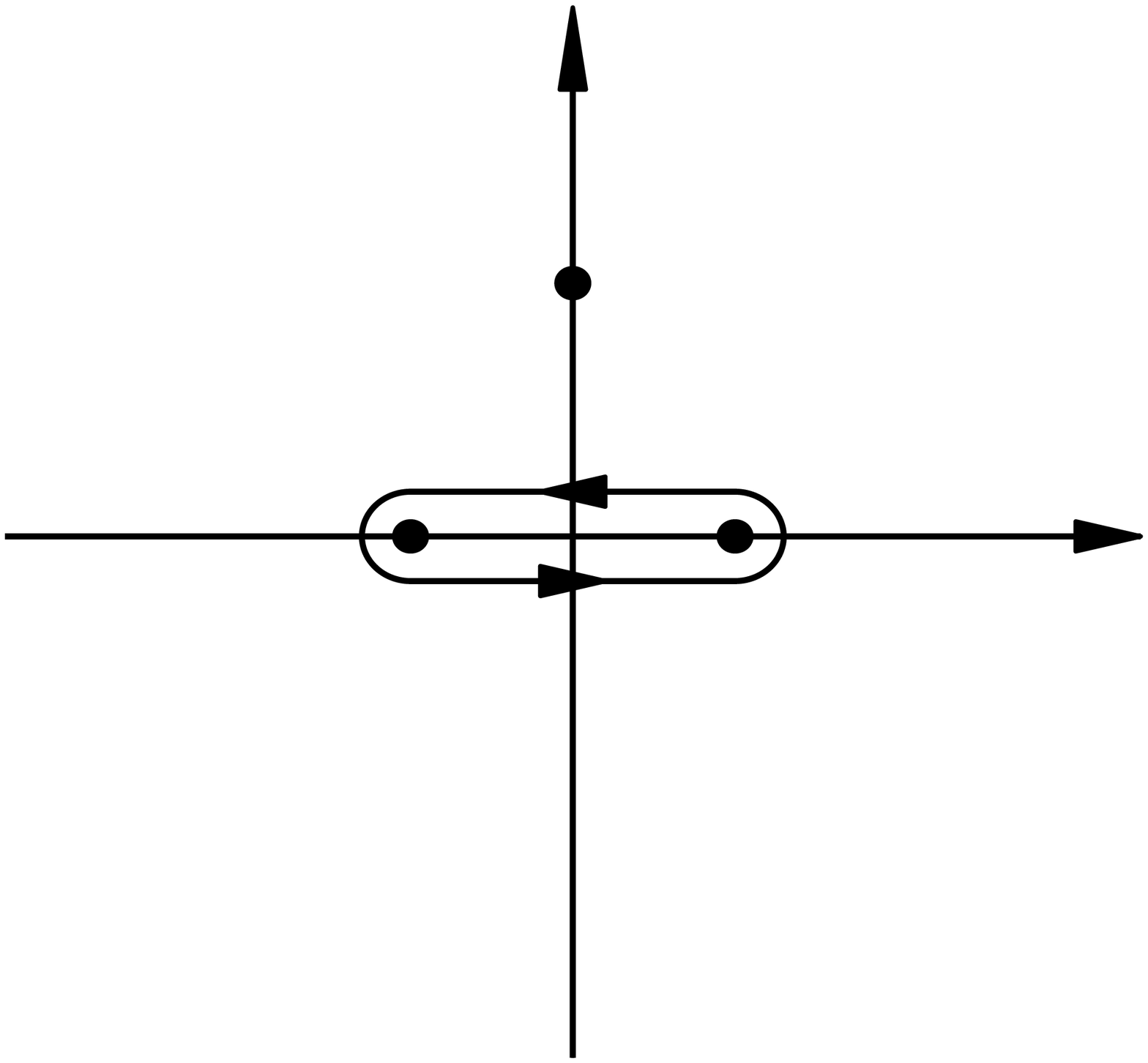}}
\put(10.2,3){$\mbox{Re}\, x$}
\put(7,5.3){$\mbox{Im}\, x$}
\put(6.8,2){$-1$} \put(8.7,2){$1$}
\put(8.2,3.95){$\frac{iK}{2m}$}
\end{picture}
\caption{The contour ${\cal C}$ in the integrals (\ref{1Loop_Integral_With_C}) and (\ref{2Loop_Integral_With_C}).}\label{Figure_Contour}
\end{figure}

\noindent
The contour ${\cal C}$ is presented in Fig. \ref{Figure_Contour}. To calculate this integral it is necessary to find the residues at the points $x=\infty$ and $x=iK/2m$. The result written as an integral over $z\equiv K^2/\Lambda^2$ has the form

\begin{equation}
\ln Z_m = \frac{\alpha_0}{\pi} \int\limits_0^\infty dz\,\frac{\Lambda^2}{4m^2 R(z)}\Big(1 - \sqrt{1+\frac{4m^2}{z\Lambda^2}} \Big) + O(\alpha_0^2).
\end{equation}

\noindent
It is convenient to introduce the new variable

\begin{equation}
\eta \equiv \frac{a}{4}\exp\Big\{-\frac{2z}{a}+\frac{2z}{a}\sqrt{1+\frac{a}{z}} -1 + \ln\Big(1+\frac{2z}{a}+2\sqrt{\frac{z}{a}+\frac{z^2}{a^2}}\Big)\Big\},
\end{equation}

\noindent
where $a\equiv 4m^2/\Lambda^2$, such that

\begin{equation}
\frac{d\eta}{2\eta} = \frac{dz}{a}\Big(-1+\sqrt{1+\frac{a}{z}}\Big).
\end{equation}

\noindent
Note that $z = 0$ and $z\to \infty$ correspond to $\eta=\exp(-1)a/4$ and $\eta\to \infty$, respectively. Therefore the integral under consideration can be rewritten as

\begin{equation}
\ln Z_m = -\frac{\alpha_0}{2\pi} \int\limits_{\exp(-1)a/4}^\infty \frac{d\eta}{\eta\, R\big(z(\eta)\big)} + O(\alpha_0^2).
\end{equation}

\noindent
This integral diverges in the limit $a\to 0$, when $z(\eta) = \eta + O(a)$. However, if the function $R^{-1}(z)$ is expanded in powers of $a$, then only the leading term will produce a divergent integral, while the other terms are given by convergent integrals, which vanish in the limit $a\to 0$. Therefore, omitting the terms suppressed by powers of $m^2/\Lambda^2$ we obtain

\begin{equation}
\ln Z_m = -\frac{\alpha_0}{2\pi} \int\limits_{\exp(-1) m^2/\Lambda^2}^\infty \frac{d\eta}{\eta\, R(\eta)} + O(\alpha_0^2).
\end{equation}

\noindent
Integrating by parts, it is possible to extract the divergent part of the remaining integral,

\begin{eqnarray}
&& \int\limits_{\exp(-1)m^2/\Lambda^2}^\infty d\eta\,\frac{1}{\eta R(\eta)} = \frac{1}{R(\eta)} \ln \eta\Big|_{\exp(-1)m^2/\Lambda^2}^\infty - \int\limits_{\exp(-1)m^2/\Lambda^2}^\infty d\eta\,\ln \eta\, \frac{d}{d\eta}\Big(\frac{1}{R(\eta)}\Big)\nonumber\\
&& = 2 \ln \frac{\Lambda}{m} +1 - A + O(m^2/\Lambda^2),
\end{eqnarray}

\noindent
where the constant $A$ is given by the equation

\begin{equation}
A \equiv \int\limits_{0}^\infty d\eta\,\ln \eta\, \frac{d}{d\eta}\Big(\frac{1}{R(\eta)}\Big),
\end{equation}

\noindent
which is equivalent to Eq. (\ref{A_Definition}). Thus, in the one-loop approximation

\begin{equation}\label{Z_M_One-Loop_Result}
\ln Z_m = -\frac{\alpha_0}{\pi} \Big(\ln\frac{\Lambda}{m}+\frac{1}{2} -\frac{A}{2}\Big) + O(\alpha_0^2).
\end{equation}

\subsection{Two-loop contribution}
\hspace*{\parindent}

To find the two-loop contribution to the renormalization constant $Z_m$, it is necessary to calculate the integrals $I_1$ --- $I_5$ in Eq. (\ref{I_One-Loop}).

The integral $I_1$ is convergent in the ultraviolet region, but diverges in the infrared one. That is why it is necessary to regularize it by introducing a small photon mass $\kappa$,

\begin{equation}
\left. I_1 = \frac{m_0^2-m^2}{\alpha_0} \int\frac{d^4K}{(2\pi)^4} \frac{8\pi}{(K^2 +\kappa^2) R_K ((P+K)^2+m^2)^2}\right|_{P^2=-m^2}.
\end{equation}

\noindent
This integral is convergent, so that it is possible to take the limit $\Lambda\to \infty$ omitting terms suppressed by powers of $\Lambda^{-1}$. Therefore, the function $R_K$ in the considered expression can be replaced by 1. The resulting integral can be calculated in the four-dimensional spherical coordinates. After the substitution $x=\cos\theta_3$ it takes the form

\begin{equation}\label{2Loop_Integral_With_C}
I_1 = \frac{m_0^2-m^2}{\alpha_0} \int\limits_0^\infty dK \frac{K}{\pi^2 (K^2+\kappa^2)} \oint\limits_{\cal C} dx \frac{\sqrt{1-x^2}}{(K+2imx)^2},
\end{equation}

\noindent
where the contour ${\cal C}$ is presented in Fig. \ref{Figure_Contour}. The integral over $x$ can be found by calculating the residues at the points $x=\infty$ and $x=iK/2m$,

\begin{eqnarray}
&& I_1 = \frac{m_0^2-m^2}{2m^2 \alpha_0} \int\limits_0^\infty dK \frac{K}{\pi (K^2+\kappa^2)} \Big(1-\frac{1}{\sqrt{1+4m^2/K^2}}\Big)\nonumber\\
&&\qquad\qquad = \frac{m_0^2-m^2}{2m^2\pi \alpha_0} \int\limits_0^\infty dK \Big(\frac{K}{K^2+\kappa^2} - \frac{1}{\sqrt{K^2+4m^2}} + \frac{\kappa^2}{(K^2+\kappa^2)\sqrt{K^2+4m^2}} \Big).\qquad
\end{eqnarray}

\noindent
Taking into account that $m_0 = Z_m m$ and omitting the last term in the round brackets (which gives a convergent integral proportional to $\kappa\to 0$) this expression can be written as

\begin{equation}
\left. I_1 = \frac{Z_m^2-1}{4\pi \alpha_0} \Big(\ln (1+K^2/\kappa^2) - 2\,\mbox{arcsinh}(K/2m) \Big)\right|_0^\infty = \frac{Z_m^2-1}{2\pi \alpha_0} \ln \frac{m}{\kappa}.
\end{equation}

\noindent
After substituting the one-loop result for $Z_m$ from Eq. (\ref{Z_M_One-Loop_Result}) with the considered accuracy the integral $I_1$ takes the form

\begin{equation}\label{I1_Result}
I_1 = -\frac{1}{\pi^2} \Big(\ln\frac{\Lambda}{m}+\frac{1}{2} -\frac{A}{2}\Big) \ln \frac{m}{\kappa}.
\end{equation}

The integral $I_2$ is convergent and does not contain infrared divergences. This implies that (due to the condition $P^2=-m^2$) this integral is equal to a finite number  and does not contribute to $\gamma_m$ in the considered approximation. Below we will omit such terms.

The integral $I_3$ diverges in the infrared region and should be regularized by introducing the small photon mass $\kappa$,

\begin{eqnarray}
&& I_3 \equiv 128\pi^2 \int\frac{d^4K}{(2\pi)^4} \frac{d^4L}{(2\pi)^4} \frac{1}{(K^2+\kappa^2) R_K (L^2+\kappa^2) R_L}\nonumber\\
&& \qquad\qquad\qquad\qquad\qquad\qquad
\left. \times \frac{m^2}{((P+K)^2+m^2)^2 ((P+K+L)^2+m^2)}\right|_{P^2=-m^2}.\qquad
\end{eqnarray}

\noindent
This expression can be equivalently rewritten as

\begin{eqnarray}\label{I3}
&& \ I_3 = 128\pi^2 \int\frac{d^4K}{(2\pi)^4}\frac{d^4L}{(2\pi)^4} \frac{m^2}{(K^2+\kappa^2) R_K (L^2+\kappa^2) R_{L}((P+K)^2+m^2)^2((P+L)^2+m^2)}\qquad\nonumber\\
&& \left.\times\left(1 + \frac{(P+L)^2-(P+K+L)^2}{(P+K+L)^2+m^2}\right)\right|_{P^2=-m^2}\equiv I_3' + I_3''.
\end{eqnarray}

\noindent
The integral $I_3'$ (corresponding to 1 in the round brackets) can be presented as a product of two integrals which have already been calculated above,

\begin{eqnarray}
&&\hspace*{-8mm} I_3' = 128\pi^2 \int\frac{d^4K}{(2\pi)^4} \frac{m^2}{(K^2+\kappa^2)R_K ((P+K)^2+m^2)^2} \int\frac{d^4L}{(2\pi)^4}\frac{1}{(L^2+\kappa^2) R_L((P+L)^2+m^2)}\nonumber\\
&&\hspace*{-8mm} = \frac{1}{2\pi^2} \ln\frac{m}{\kappa}\Big(2\ln\frac{\Lambda}{m}+1-A\Big).
\end{eqnarray}

\noindent The expression for $I_3''$ (which is obtained from the second term in the round brackets in Eq. (\ref{I3})) is not divergent in the infrared region, so that it is possible to set $\kappa$ to 0,

\begin{eqnarray}
&&\hspace*{-8mm} I_3''\Big|_{\kappa=0} = - 128\pi^2\int\frac{d^4K}{(2\pi)^4}\frac{d^4L}{(2\pi)^4}\frac{1}{K^2 R_K L^2 R_L}\nonumber\\
&&\hspace*{-8mm} \qquad\qquad\qquad\qquad
\left. \times \frac{m^2 K^\mu (K+2P+2L)_\mu}{((P+K)^2+m^2)^2((P+L)^2+m^2)((P+K+L)^2+m^2)} \right|_{P^2=-m^2}.\qquad
\end{eqnarray}

\noindent
It is also not divergent in the ultraviolet region. Therefore, it is a finite number, which does not contribute to the two-loop mass anomalous dimension. This implies that

\begin{equation}\label{I3_Result}
I_3 = \frac{1}{2\pi^2} \ln\frac{m}{\kappa}\Big(2\ln\frac{\Lambda}{m}+1-A\Big) + \mbox{a finite constant}.
\end{equation}

After some transformations the expression $I_4 - \big(I_{\mbox{\scriptsize one-loop}}\big)^2/2$ can be rewritten as

\begin{eqnarray}\label{I4_Modified}
&& I_4 - \frac{1}{2}\big(I_{\mbox{\scriptsize one-loop}}\big)^2 = - 64\pi^2 \int\frac{d^4K}{(2\pi)^4}\frac{d^4L}{(2\pi)^4}\frac{1}{K^2 R_K L^2 R_L} \nonumber\\
&&\qquad\qquad\qquad\left. \times \frac{L_\mu K^\mu}{((P+K)^2+m^2)((P+L)^2+m^2)((P+K+L)^2+m^2)}\right|_{P^2=-m^2}.\qquad
\end{eqnarray}

\noindent
This integral is convergent in the infrared region. Therefore, it depends on $\Lambda/m$ and can be presented as

\begin{equation}\label{I4_Structure}
I_4 - \frac{1}{2}\big(I_{\mbox{\scriptsize one-loop}}\big)^2 = f_2\ln^2\frac{\Lambda}{m} + f_1 \ln \frac{\Lambda}{m} + f_0 + \mbox{terms proportional to $m/\Lambda$}.
\end{equation}

\noindent
Let us calculate the derivative of the integral (\ref{I4_Modified}) with respect to $\ln\Lambda$ in the case $m=0$,

\begin{equation}
\left.\frac{d}{d\ln\Lambda}\Big(I_4 - \frac{1}{2}\big(I_{\mbox{\scriptsize one-loop}}\big)^2\Big)\right|_{m\to 0} = - 64\pi^2 \frac{d}{d\ln\Lambda}\int\frac{d^4K}{(2\pi)^4}\frac{d^4L}{(2\pi)^4}\frac{L_\mu K^\mu}{K^4 R_K L^4 R_L (K+L)^2}.
\end{equation}

\noindent
Following Ref. \cite{Soloshenko:2003nc}, this integral can be presented as

\begin{eqnarray}
&&\frac{1}{2\pi^2} \left[ \int\limits_0^{\infty}\frac{dK}{K^3} \int\limits_0^K dL\, L \frac{d}{d\ln\Lambda}\Big(\frac{1}{R_K R_L}\Big) + \int\limits_0^{\infty} dK\, K \int\limits_K^{\infty}\frac{dL}{L^3}\frac{d}{d\ln\Lambda} \Big(\frac{1}{R_K R_L}\Big)\right]\qquad\nonumber\\
&& = \frac{1}{\pi^2} \int\limits_0^{\infty}\frac{dK}{K^3} \int_0^K dL\, L \frac{d}{d\ln\Lambda}\Big(\frac{1}{R_K R_L}\Big).
\end{eqnarray}

\noindent
After the substitution $L=\rho K$ in the last integral, this expression can be rewritten in the form\footnote{This result agrees with the calculation of Ref. \cite{Soloshenko:2003nc} carried out for the particular case $R(x) = 1+x^n$.}

\begin{equation}\label{Integral_For_All_R}
\frac{1}{\pi^2} \int\limits_0^{\infty}\frac{dK}{K} \int\limits_0^1 d\rho\, \rho \frac{d}{d\ln\Lambda}\Big(\frac{1}{R_K R_{\rho K}}\Big) = - \frac{1}{\pi^2} \int\limits_0^1 d\rho\, \rho \int\limits_0^{\infty} d\ln K\, \frac{d}{d\ln K}\Big(\frac{1}{R_K R_{\rho K}}\Big) = \frac{1}{2\pi^2}.
\end{equation}

\noindent
This implies that in Eq. (\ref{I4_Structure}) $f_2 = 0$ and $f_1 = 1/2\pi^2$. Therefore, omitting terms proportional to $m/\Lambda$, we obtain

\begin{equation}\label{I4_Result}
I_4 - \frac{1}{2}\big(I_{\mbox{\scriptsize one-loop}}\big)^2 = \frac{1}{2\pi^2} \ln \frac{\Lambda}{m} + \mbox{a finite constant}.
\end{equation}

The remaining integral $I_5$ can be calculated using the equation

\begin{eqnarray}
&& \int \frac{d^4L}{(2\pi)^4} \frac{1}{(L^2+m^2)((K+L)^2+m^2)} + \sum\limits_{I=1}^n (-1)^{P_I} \int \frac{d^4L}{(2\pi)^4} \frac{1}{(L^2+M_I^2)((K+L)^2+M_I^2)}\nonumber\\
&& = J(K/m) + \sum\limits_{I=1}^n (-1)^{P_I} J(K/M_I),\vphantom{\frac{1}{2}}
\end{eqnarray}

\noindent
where

\begin{eqnarray}
&& J(K/M) \equiv -\frac{1}{8\pi^2}\Big(\ln\frac{M}{K} + \sqrt{1+\frac{4M^2}{K^2}}  \mbox{arctanh}\sqrt{\frac{K^2}{K^2+4M^2}}\Big)  \qquad\nonumber\\
&&\vphantom{1} \nonumber\\
&&\qquad\qquad\qquad\qquad\qquad\qquad\qquad\qquad\quad \approx \left\{\begin{array}{l}
{\displaystyle \frac{1}{4\pi^2} \frac{M^2}{K^2} \ln \frac{M}{K} \qquad\quad \mbox{if}\quad K\to \infty}\\
\vphantom{1}\\
{\displaystyle -\frac{1}{8\pi^2}\Big(\ln\frac{M}{K}+1\Big) \quad \mbox{if}\quad K\to 0}
\end{array}
\right.,\qquad\quad
\end{eqnarray}

\noindent
see, e.g., Ref. \cite{Soloshenko:2003nc}. This implies that the integral

\begin{equation}\label{Pauli-Villars_Integral}
\left. 64\pi^2 N_f \int\frac{d^4K}{(2\pi)^4}\frac{1}{K^2 R_K^2((P+K)^2+m^2)} J(K/m)\right|_{P^2=-m^2}
\end{equation}

\noindent
is convergent in both ultraviolet and infrared regions. Therefore, it is equal to a finite constant, and only the terms with the Pauli--Villars masses $M_I$ nontrivially contribute to the divergent part of the integral $I_5$. To calculate them, let us consider the expression

\begin{equation}
\left. 64\pi^2 N_f \frac{d}{d\ln\Lambda}\int\frac{d^4K}{(2\pi)^4}\frac{1}{K^2 R_K^2((P+K)^2+m^2)} J(K/M) \right|_{P^2=-m^2} \equiv J_1 + J_2 + J_3,
\end{equation}

\noindent
where $M = a\Lambda$ with $a$ being a finite constant and

\begin{eqnarray}
&&\hspace*{-12mm} J_1 \equiv \left. - 8 N_f \int\frac{d^4K}{(2\pi)^4}\frac{1}{K^2 R_K^2((P+K)^2+m^2)} \right|_{P^2=-m^2};\qquad\\
&&\hspace*{-12mm} J_2 \equiv \left. - 8 N_f \int\frac{d^4K}{(2\pi)^4}\frac{1}{K^2 ((P+K)^2+m^2)} \frac{d}{d\ln\Lambda}\Big(\frac{1}{R_K^2}\Big) \ln\frac{\Lambda}{K}\right|_{P^2=-m^2};\\
&&\hspace*{-12mm} J_3 \equiv \left. 64\pi^2 N_f \int\frac{d^4K}{(2\pi)^4}\frac{1}{K^2 ((P+K)^2+m^2)} \frac{d}{d\ln\Lambda}\Big[\frac{1}{R_K^2}\Big(J(K/M) + \frac{1}{8\pi^2} \ln\frac{\Lambda}{K}\Big)\Big]\right|_{P^2=-m^2}.
\end{eqnarray}

\noindent
Repeating the calculation of Appendix \ref{Appendix_One-Loop} we obtain

\begin{equation}
J_1  = - \frac{N_f}{\pi^2} \Big(\ln\frac{\Lambda}{m} + \frac{1}{2}\Big) - J_2 + O\Big(\frac{m^2}{\Lambda^2}\Big).
\end{equation}

\noindent
To find the integral $J_3$, we note that the derivative of the function in the square brackets with respect to $\ln\Lambda$ is equal to the one with respect to $\ln K$ multiplied by $(-1)$. Therefore,

\begin{eqnarray}
&& J_3\Big|_{m\to 0} = - 8 N_f \int\limits_{0}^\infty \frac{dK}{K}\, \frac{d}{d \ln K}\Big[\frac{1}{R_K^2}\Big(J(K/M) + \frac{1}{8\pi^2} \ln\frac{\Lambda}{K}\Big)\Big]\nonumber\\
&&\left. = - 8 N_f \Big[\frac{1}{R_K^2}\Big(J(K/M) + \frac{1}{8\pi^2} \ln\frac{\Lambda}{K}\Big)\Big]\right|_0^\infty = -\frac{N_f}{\pi^2} \big(1+\ln a\big).
\end{eqnarray}

\noindent
This implies that the expression (\ref{Pauli-Villars_Integral}) can be written as

\begin{equation}
\frac{N_f}{\pi^2} \Big[ - \frac{1}{2}\ln^2\frac{\Lambda}{m} - \ln\frac{\Lambda}{m}\Big(\ln a + \frac{3}{2} \Big)\Big] + O(1).
\end{equation}

\noindent
Consequently, the integral $I_5$ takes the form

\begin{equation}\label{I5_Result}
I_5 = \frac{N_f}{\pi^2}\Big[\, \frac{1}{2}\ln^2\frac{\Lambda}{m} + \ln\frac{\Lambda}{m}\Big(\sum_{I=1}^{n}c_I\ln a_I+\frac{3}{2}\Big) + \mbox{a finite constant}\Big],
\end{equation}

\noindent
where $c_I=(-1)^{P_I+1}$ and terms vanishing in the limit $\Lambda\to \infty$ were omitted.

Collecting the results (\ref{Z_M_One-Loop_Result}), (\ref{I1_Result}), (\ref{I3_Result}), (\ref{I4_Result}), and (\ref{I5_Result}) we obtain that the two-loop mass renormalization constant is given by the expression (\ref{MassRenorm}),

\begin{eqnarray}\label{Z_M_Result}
&&\ln Z_m = - \frac{\alpha_0}{\pi}\Big(\ln\frac{\Lambda}{m}+\frac{1}{2}-\frac{A}{2}\Big) + \frac{\alpha_0^2}{\pi^2}\Big[\, \frac{N_f}{2}\ln^2\frac{\Lambda}{m}+ \ln\frac{\Lambda}{m}\Big(N_f \sum_{I=1}^{n}c_I\ln a_I+\frac{3N_f+1}{2}\Big)\qquad\nonumber\\
&& +\mbox{a finite constant}\Big] + O(\alpha_0^3),
\end{eqnarray}

\noindent
and does not contain infrared divergences.


\begin{thebibliography}{100}

\bibitem{Grozin:2005yg}
  A.~Grozin,
  ``Lectures on QED and QCD,'' Lectures at 3rd Dubna International Advanced School of Theoretical Physics	
  29 Jan - 6 Feb 2005. Dubna, Russia,
  [hep-ph/0508242].

\bibitem{Novikov:1983uc}
  V.~A.~Novikov, M.~A.~Shifman, A.~I.~Vainshtein and V.~I.~Zakharov,
  Nucl.\ Phys.\ B {\bf 229} (1983) 381.

\bibitem{Jones:1983ip}
  D.~R.~T.~Jones,
  Phys.\ Lett.\  {\bf 123B} (1983) 45.

\bibitem{Novikov:1985rd}
  V.~A.~Novikov, M.~A.~Shifman, A.~I.~Vainshtein and V.~I.~Zakharov,
  Phys.\ Lett.\  {\bf 166B} (1986) 329
   [Sov.\ J.\ Nucl.\ Phys.\  {\bf 43} (1986) 294]
   [Yad.\ Fiz.\  {\bf 43} (1986) 459].

\bibitem{Shifman:1986zi}
  M.~A.~Shifman and A.~I.~Vainshtein,
  Nucl.\ Phys.\ B {\bf 277} (1986) 456
   [Sov.\ Phys.\ JETP {\bf 64} (1986) 428]
   [Zh.\ Eksp.\ Teor.\ Fiz.\  {\bf 91} (1986) 723].

\bibitem{Vainshtein:1986ja}
  A.~I.~Vainshtein, V.~I.~Zakharov and M.~A.~Shifman,
  JETP Lett.\  {\bf 42} (1985) 224
   [Pisma Zh.\ Eksp.\ Teor.\ Fiz.\  {\bf 42} (1985) 182].

\bibitem{Shifman:1985fi}
  M.~A.~Shifman, A.~I.~Vainshtein and V.~I.~Zakharov,
  Phys.\ Lett.\  {\bf 166B} (1986) 334.

\bibitem{Kutasov:2004xu}
  D.~Kutasov and A.~Schwimmer,
  Nucl.\ Phys.\ B {\bf 702} (2004) 369.

\bibitem{Kataev:2014gxa}
  A.~L.~Kataev and K.~V.~Stepanyantz,
  Theor.\ Math.\ Phys.\  {\bf 181} (2014) 1531.

\bibitem{Goriachuk:2018cac}
  I.~O.~Goriachuk, A.~L.~Kataev and K.~V.~Stepanyantz,
  Phys.\ Lett.\ B {\bf 785} (2018) 561.

\bibitem{Stueckelberg:1953dz}
  E.~G.~Stueckelberg and A.~Petermann,
  Helv.\ Phys.\ Acta {\bf 26} (1953) 499.

\bibitem{GellMann:1954fq}
  M.~Gell-Mann and F.~E.~Low,
  Phys.\ Rev.\  {\bf 95} (1954) 1300.

\bibitem{Bogolyubov:1956gh}
  N.~N.~Bogolyubov and D.~V.~Shirkov,
  Nuovo Cim.\  {\bf 3} (1956) 845.

\bibitem{Bogolyubov:1980nc}
  N.~N.~Bogolyubov and D.~V.~Shirkov,
  ``Introduction To The Theory Of Quantized Fields,''
  Intersci.\ Monogr.\ Phys.\ Astron.\  {\bf 3} (1959) 1 [Moscow: Nauka, the fourth edition, (1984) 416 p, In Russian].

\bibitem{Kataev:2013eta}
  A.~L.~Kataev and K.~V.~Stepanyantz,
  Nucl.\ Phys.\ B {\bf 875} (2013) 459.

\bibitem{Kataev:2013csa}
  A.~L.~Kataev and K.~V.~Stepanyantz,
  Phys.\ Lett.\ B {\bf 730} (2014) 184.

\bibitem{Slavnov:1971aw}
  A.~A.~Slavnov,
  Nucl.\ Phys.\ B {\bf 31} (1971) 301.

\bibitem{Slavnov:1972sq}
  A.~A.~Slavnov,
  Theor.Math.Phys. {\bf 13} (1972) 1064
   [Teor.\ Mat.\ Fiz.\  {\bf 13} (1972) 174].

\bibitem{Krivoshchekov:1978xg}
  V.~K.~Krivoshchekov,
  Theor.\ Math.\ Phys.\ {\bf 36} (1978) 745
 [Teor.\ Mat.\ Fiz.\  {\bf 36} (1978) 291].

\bibitem{West:1985jx}
  P.~C.~West,
  Nucl.\ Phys.\ B {\bf 268} (1986) 113.

\bibitem{Aleshin:2016yvj}
  S.~S.~Aleshin, A.~E.~Kazantsev, M.~B.~Skoptsov and K.~V.~Stepanyantz,
  JHEP {\bf 1605} (2016) 014.

\bibitem{Nartsev:2016mvn}
  I.~V.~Nartsev and K.~V.~Stepanyantz,
  JETP Lett.\  {\bf 105} (2017) no.2,  69.

\bibitem{Hisano:1997ua}
  J.~Hisano and M.~A.~Shifman,
  Phys.\ Rev.\ D {\bf 56} (1997) 5475.

\bibitem{Jack:1997pa}
  I.~Jack and D.~R.~T.~Jones,
  Phys.\ Lett.\ B {\bf 415} (1997) 383.

\bibitem{Avdeev:1997vx}
  L.~V.~Avdeev, D.~I.~Kazakov and I.~N.~Kondrashuk,
  Nucl.\ Phys.\ B {\bf 510} (1998) 289.

\bibitem{Kataev:2017qvk}
  A.~L.~Kataev, A.~E.~Kazantsev and K.~V.~Stepanyantz,
  Nucl.\ Phys.\ B {\bf 926} (2018) 295.

\bibitem{Stepanyantz:2011jy}
  K.~V.~Stepanyantz,
  Nucl.\ Phys.\ B {\bf 852} (2011) 71.

\bibitem{Stepanyantz:2014ima}
  K.~V.~Stepanyantz,
  JHEP {\bf 1408} (2014) 096.

\bibitem{Nartsev:2016nym}
  I.~V.~Nartsev and K.~V.~Stepanyantz,
  JHEP {\bf 1704} (2017) 047.

\bibitem{Shifman:2014cya}
  M.~Shifman and K.~Stepanyantz,
  Phys.\ Rev.\ Lett.\  {\bf 114} (2015) no.5,  051601.

\bibitem{Shifman:2015doa}
  M.~Shifman and K.~V.~Stepanyantz,
  Phys.\ Rev.\ D {\bf 91} (2015) 105008.

\bibitem{Stepanyantz:2016gtk}
  K.~V.~Stepanyantz,
  Nucl.\ Phys.\ B {\bf 909} (2016) 316.

\bibitem{Shakhmanov:2017soc}
  V.~Y.~Shakhmanov and K.~V.~Stepanyantz,
  Nucl.\ Phys.\ B {\bf 920} (2017) 345.

\bibitem{Kazantsev:2018nbl}
  A.~E.~Kazantsev, V.~Y.~Shakhmanov and K.~V.~Stepanyantz,
  JHEP {\bf 1804} (2018) 130.

\bibitem{Jack:1996vg}
  I.~Jack, D.~R.~T.~Jones and C.~G.~North,
  Phys.\ Lett.\ B {\bf 386} (1996) 138.

\bibitem{Harlander:2006xq}
  R.~V.~Harlander, D.~R.~T.~Jones, P.~Kant, L.~Mihaila and M.~Steinhauser,
  JHEP {\bf 0612} (2006) 024.

\bibitem{Jack:1996cn}
  I.~Jack, D.~R.~T.~Jones and C.~G.~North,
  Nucl.\ Phys.\ B {\bf 486} (1997) 479.

\bibitem{Jack:1998uj}
  I.~Jack, D.~R.~T.~Jones and A.~Pickering,
  Phys.\ Lett.\ B {\bf 435} (1998) 61.

\bibitem{Aleshin:2016rrr}
  S.~S.~Aleshin, I.~O.~Goriachuk, A.~L.~Kataev and K.~V.~Stepanyantz,
  Phys.\ Lett.\ B {\bf 764} (2017) 222.

\bibitem{Smilga:2004zr}
  A.~V.~Smilga and A.~Vainshtein,
  Nucl.\ Phys.\ B {\bf 704} (2005) 445.

\bibitem{Smilga_Private}
 A.~V.~Smilga, private communication, 2017.

\bibitem{Slavnov:1977zf}
  A.~A.~Slavnov,
  Theor.\ Math.\ Phys. {\bf 33} (1977) 977 [Teor.\ Mat.\ Fiz.\  {\bf 33} (1977) 210].

\bibitem{Kazantsev:2014yna}
  A.~E.~Kazantsev and K.~V.~Stepanyantz,
  J.\ Exp.\ Theor.\ Phys.\  {\bf 120} (2015) no.4,  618
   [Zh.\ Eksp.\ Teor.\ Fiz.\  {\bf 147} (2015) no.4,  714].

\bibitem{Grisaru:1979wc}
  M.~T.~Grisaru, W.~Siegel and M.~Rocek,
  Nucl.\ Phys.\ B {\bf 159} (1979) 429.

\bibitem{Wess:1974jb}
  J.~Wess and B.~Zumino,
  Nucl.\ Phys.\ B {\bf 78} (1974) 1.

\bibitem{Goity:1983aw}
  J.~Goity, T.~Kugo and R.~D.~Peccei,
  Phys.\ Rev.\ D {\bf 29} (1984) 2412.

\bibitem{Soloshenko:2003nc}
  A.~A.~Soloshenko and K.~V.~Stepanyantz,
  Theor.\ Math.\ Phys.\  {\bf 140} (2004) 1264
   [Teor.\ Mat.\ Fiz.\  {\bf 140} (2004) 437].

\bibitem{Pimenov:2009hv}
  A.~B.~Pimenov, E.~S.~Shevtsova and K.~V.~Stepanyantz,
  Phys.\ Lett.\ B {\bf 686} (2010) 293.

\bibitem{Stepanyantz:2011bz}
  K.~V.~Stepanyantz,
  ``Factorization of integrals defining the two-loop $\beta$-function for the general renormalizable N=1 SYM theory, regularized by the higher covariant derivatives, into integrals of double total derivatives,''
  arXiv:1108.1491 [hep-th].

\bibitem{Buchbinder:2015eva}
  I.~L.~Buchbinder, N.~G.~Pletnev and K.~V.~Stepanyantz,
  Phys.\ Lett.\ B {\bf 751} (2015) 434.

\bibitem{Stepanyantz:2012zz}
  K.~V.~Stepanyantz,
  J.\ Phys.\ Conf.\ Ser.\  {\bf 343} (2012) 012115.

\bibitem{Broadhurst:1991fy}
  D.~J.~Broadhurst, N.~Gray and K.~Schilcher,
  Z.\ Phys.\ C {\bf 52} (1991) 111.

\bibitem{Gnendiger:2017pys}
  C.~Gnendiger {\it et al.},
  Eur.\ Phys.\ J.\ C {\bf 77} (2017) no.7,  471.

\bibitem{Kataev:2013vua}
  A.~L.~Kataev,
  JHEP {\bf 1402} (2014) 092.

\bibitem{Vladimirov:1979my}
  A.~A.~Vladimirov,
  Sov.\ J.\ Nucl.\ Phys.\  {\bf 31} (1980) 558
   [Yad.\ Fiz.\  {\bf 31} (1980) 1083].

\bibitem{Broadhurst:1992za}
  D.~J.~Broadhurst, A.~L.~Kataev and O.~V.~Tarasov,
  Phys.\ Lett.\ B {\bf 298} (1993) 445.

\bibitem{Soloshenko:2002np}
  A.~Soloshenko and K.~Stepanyantz,
  ``Two loop renormalization of N=1 supersymmetric electrodynamics, regularized by higher derivatives,''
  hep-th/0203118.

\bibitem{Soloshenko:2003sx}
  A.~A.~Soloshenko and K.~V.~Stepanyantz,
  Theor.\ Math.\ Phys.\  {\bf 134} (2003) 377
   [Teor.\ Mat.\ Fiz.\  {\bf 134} (2003) 430].

\end{thebibliography}
\end{document}